# Metamaterial optical diodes for linearly and circularly polarized light

**E. Plum, V. A. Fedotov and N. I. Zheludev**
*Optoelectronics Research Centre, University of Southampton, SO17 1BJ, UK*
*erp@orc.soton.ac.uk*

**Abstract:** The total intensity of light transmitted at non-normal incidence thorough planar metamaterials can be different for forward and backward propagation. For metamaterial patterns of different symmetries we observe this effect for circularly or linearly polarized light.
**OCIS codes:** (160.3918) Metamaterials; (160.4760) Optical properties; (240.5445) Polarization-selective devices

We demonstrate a new family of metamaterial devices which exhibits different levels of total transmission for opposite propagation directions. The optical diode functionality does not violate Lorentz reciprocity [1] and arises from different rates of polarization conversion for waves entering the device from opposite sides. Two families of metamaterials, exhibiting the optical diode functionality for linearly and circularly polarized light correspondingly, have been identified and tested in the microwave part of the spectrum.

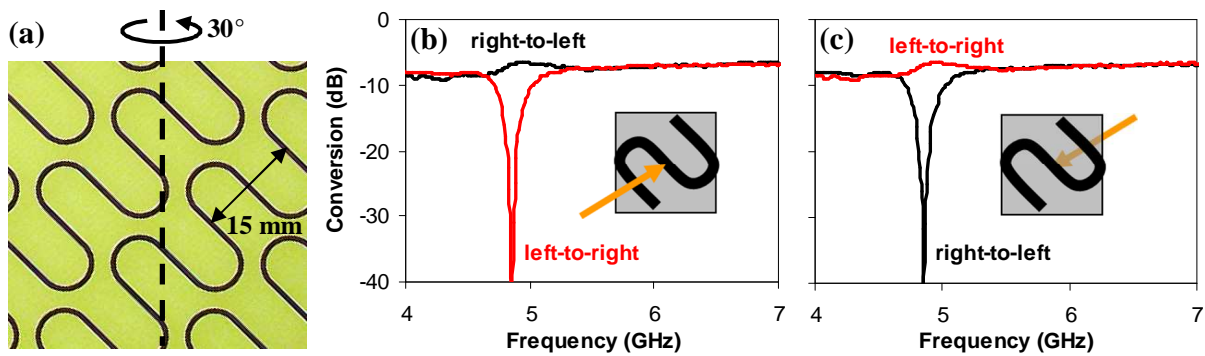

Fig. 1. **Optical diode for circularly polarized waves.** (a) Achiral planar metamaterial with twofold rotational symmetry (wallpaper symmetry group *pmg*). The structure consists of meandering metal wires sandwiched in between lossy dielectric substrates and was tilted around the axis indicated by the dashed line. At 30° oblique incidence, the metamaterial exhibits reversed left-to-right and right-to-left polarization conversion efficiencies for circularly polarized waves incident on its (b) front and (c) back.

Directionally asymmetric total transmission of circularly polarized waves has been previously reported for normal incidence onto lossy, anisotropic, 2D-chiral patterns (wallpaper symmetry groups: *p1, p2*) [2, 3]. We show that diode-like behavior for circularly polarized waves can be observed at oblique incidence onto any lossy periodic pattern. Here lossy, non-chiral, planar metamaterials with twofold rotational symmetry (wallpaper symmetry groups: *pmm, pmg, pgg, cmm, p4m, p4g, p6m*) are of particular interest, as they exhibit diode-like behavior at oblique incidence while circular dichroism is absent. For such structures, the effect occurs when – starting from normal incidence – the metamaterial is tilted around any axis that is neither parallel nor perpendicular to a line of mirror symmetry, see Fig. 1 (a). The directional asymmetry in total transmission results from different circular polarization conversion efficiencies (e.g., left-to-right) for opposite directions of propagation, see Fig. 1 (b) and (c). At about 4.8 GHz, the example of an optical diode shown here is more transparent for right-handed circularly polarized waves incident on its front than its back.

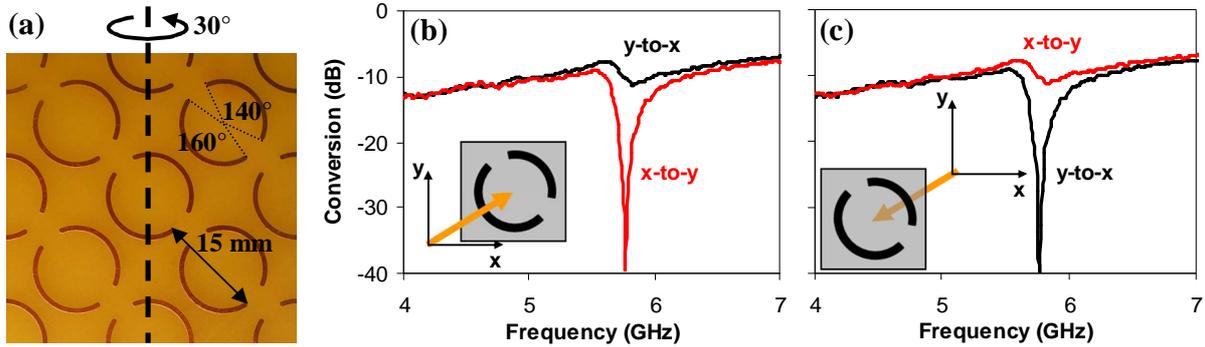

Fig. 2. **Optical diode for linearly polarized waves.** (a) Planar metamaterial without twofold rotational symmetry (wallpaper symmetry group *pm*). The structure consists of asymmetrically split wire rings supported by a lossy dielectric substrate and it was tilted around the axis indicated by the dashed line. At 30° oblique incidence, the metamaterial exhibits reversed x-to-y and y-to-x polarization conversion levels for linearly polarized waves incident on its (b) front and (c) back.

Directionally asymmetric transmission of linearly polarized waves has recently been demonstrated for a low symmetry three-dimensional metamaterial structure [4]. We show for the first time that diode-like behavior for linearly polarized waves also occurs in planar metamaterials. This phenomenon can be observed at oblique incidence onto any lossy planar metamaterial without twofold rotational symmetry (wallpaper symmetry groups: *p1, p3, pm, pg, cm, p3m1, p31m*). The effect occurs when – starting from normal incidence – the metamaterial is tilted around any axis that is neither parallel nor perpendicular to a line of mirror symmetry of the metamaterial pattern, see Fig. 2 (a). In this case the effect results from directionally asymmetric linear polarization conversion efficiencies (e.g., x-to-y), see Fig. 2 (b) and (c). At about 5.8 GHz, this example of an optical diode exhibits larger total transmission of y-polarization incident on its front than its back.

The transmission characteristics of the metamaterial structures shown in Figs. 1 (a) and 2 (a) were investigated in a microwave anechoic chamber using linearly polarized broadband horn antennas (Schwarzbeck BBHA 9120D) equipped with lens concentrators and a vector network analyzer (Agilent E8364B). Transmission of circularly polarized waves was studied by transforming the measured complex transmission matrix (Jones matrix) from the linear polarization basis to the circular polarization basis.

In both cases the directional transmission asymmetries are accounted for by directionally asymmetric absorption losses. How fundamentally different the observed effects are from conventional polarization effects becomes clear when considering the associated eigenstates. An ideal planar metamaterial diode for circularly polarized waves would have a single degenerate left-handed circular eigenstate for forward-propagation and a right-handed circular eigenstate for backward propagation (or vice versa). On the other hand, an ideal linear polarization diode would have single degenerate linearly polarized eigenstates of orthogonal orientations for opposite directions of propagation.

In summary we demonstrate planar metamaterial optical diodes for linearly and circularly polarized electromagnetic waves.